\documentclass[]{aa} 

\usepackage{graphicx}

\usepackage{txfonts}
\usepackage{answers}
\begin{document}

   \title{CanariCam/GTC observations of (99942) Apophis}


   \author{J. Licandro\inst{1,2}
          \and
          T. M\"uller\inst{3}
          \and
          C. Alvarez\inst{1,2}
             \and
          V. Al\'i-Lagoa\inst{4}
          \and
          M. Delbo\inst{4}
          }
  \institute{Instituto de Astrof\'isica de Canarias (IAC),
             C/V\'ia L\'actea s/n, 38205 La Laguna, Tenerife, Spain.\\
              \email{jlicandr@iac.es}
  \and
  Departamento de Astrof\'{\i}sica, Universidad de La Laguna. 38206, La Laguna, Tenerife, Spain
  \and
 Max-Planck-Institut f\"ur extraterrestrische Physik, Postfach 1312, Giessenbachstrasse, 85741 Garching, Germany
  \and 
  UNS-CNRS-Observatoire de la C\^ote d'Azur, B.P. 4229, 06304 Nice Cedex 4, France
              }

   \date{Received September 15, 1996; accepted March 16, 1997}

 
  \abstract
  {The potentially hazardous  asteroid (PHA) (99942) Apophis is one of the most remarkable near-Earth asteroids (NEA) in terms of impact hazard. A good determination of its surface thermal inertia is very important in order to evaluate the Yarkovsky effect on its orbital evolution. }
   {We present thermal infrared observations obtained on January 29, 2013, with CanariCam mid-infrared camera/spectrograph attached to the  Gran Telescopio CANARIAS (GTC, Roque de los Muchachos Observatory, La Palma, Spain) using the Si2-8.7, Si6-12.5, and Q1-17.65 filters  with the aim of deriving Apophis' diameter ($D$), geometric albedo ($p_V$), and thermal inertia ($\Gamma$).}
   { We performed a detailed thermophysical model analysis of the GTC data combined with previously published thermal data obtained using  {\textit Herschel} {\textit Space} {\textit Observatory} PACS instrument at 70, 100, and 160 $\mu$m.}
   {The thermophysical model fit of the data favors low surface roughness solutions (within a range of roughness slope angles $rms$ between 0.1 and 0.5), and constrains the effective diameter, visible geometric albedo, and thermal inertia of Apophis to be $D_{eff} =$~380 -- 393 m, $p_V = $~0.24--0.33 (assuming absolute magnitude $H = 19.09 \pm 0.19$) and $\Gamma =$~50 -- 500 Jm$^{-2}$ s$^{-0.5}$ K$^{-1}$, respectively.}
   {}


   \maketitle
%

\section{Introduction}

The potentially hazardous  asteroid (PHA) (99942) Apophis (hereafter Apophis) is a near-Earth asteroid (NEA) that has a small but non-zero chance of impacting the Earth and it is one of the most remarkable NEAs in terms of impact hazard. With the available data, it is known that  Apophis will have an extremely close approach on April 13, 2029, at 5.7 Earth radii from the Earth's center, just below the altitude of geosynchronous Earth satellites;  by means of a statistical analysis, Farnocchia et al. \cite{Farnocchia2013}  find an impact probability greater than $10^{-6}$ for an impact in 2068. 

The computation of the orbital evolution of this object is limited by the insufficient knowledge of the role played by the non-gravitational Yarkovsky effect, which produces a steady orbital drift as a consequence of the momentum carried away by the thermal emission of the object (see, e.g., Bottke et al., \cite{Botkke2002}, Giorgini et al., \cite{Giorgini2002}). The Yarkovsky effect depends upon several poorly known parameters, such as the albedo, size, thermal inertia, and pole orientation of the object. Recently Vokrouhlick\'y et al. \cite{Vokrouhlicky} evaluated the Yarkovsky effect on the orbital evolution of Apophis. 

In addition, improved knowledge of the physical properties of Apophis is desirable for other reasons: (1)  the European Commission H2020-PROTEC-2014 funded project NEOShield-2  will base its study of NEO mitigation strategy on the case of Apophis and (2)  the possible implications would need to be addressed if an impact were to occur.

A first determination from polarimetric observations of the geometric albedo  $p_V = 0.33 \pm 0.08$ is presented in Delbo et al. \cite{Delbo2007}. They also obtained an absolute magnitude of $H = 19.7 \pm 0.4$ mag, which led to an effective diameter $D_{eff}  = 270 \pm 60$ m, slightly smaller than earlier estimates in the range of 320 to 970 m depending on the assumed albedo.  
M\"uller et al. \cite{Mueller2014} published the first far-infrared observations of Apophis using the {\em Herschel Space Observatory} PACS instrument. They obtained data at 70, 100, and 160 $\mu$m at two epochs and performed a detailed thermophysical model (TPM) analysis. They used the spin and shape model and absolute magnitude $H = 19.09 \pm 0.19$ by Pravec et al. \cite{Pravec2014} and obtained an effective diameter $D_{eff} = 375 ^{+14}_{-10}$ m, a geometric albedo in the $V$-band $p_V = 0.30 ^{+0.05}_{-0.06}$, and a thermal inertia of $\Gamma = 600 ^{+200}_{-350}$ Jm$^{-2}$s$^{-0.5}$K$^{-1}$. The albedo determinations agree very well; the difference between the $D_{eff}$ determined by M\"uller et al. and the value derived by Delbo et al. \cite{Delbo2007} from their $p_V$ determination is the result of a different value of $H$ ($H = 19.09$ and $H = 19.7,$ respectively). The Vokrouhlick\'y et al. \cite{Vokrouhlicky} results use the Pravec et al. \cite{Pravec2014} and  M\"uller et al. \cite{Mueller2014} results to evaluate the Yarkovsky effect on the orbital evolution of Apophis; in particular, they use the range of $\Gamma$ values provided by M\"uller et al. \cite{Mueller2014}.

In this paper we present thermal infrared observations obtained on January 29, 2013, with the CanariCam mid-infrared instrument attached to the  Gran Telescopio CANARIAS (GTC, Roque de los Muchachos Observatory, La Palma, Spain). Images of Apophis were obtained using three different filters (Si2-8.7, Si6-12.5, and Q1-17.65). These fluxes, obtained at wavelengths which are closer to the wavelengths  in which the Apophis thermal emission peaks than the {\em Herschel} measurements, are used together with {\em Herschel} data to better constrain the thermophysical model presented in M\"uller et al. \cite{Mueller2014}. 

The paper is organized as follows. In Sect. \ref{sec:data} we present the observations and describe the reduction process and photometry.  The TPM is described and the results are presented in Sect. \ref{sec:TPM}. Finally the discussion and conclusions are presented in Sect. \ref{sec:discussion}. 


\section{Observations and data reduction}\label{sec:data}

The observations were performed on January 29, 2013, with CanariCam (see Telesco et al. \cite{Telesco}) in imaging mode at the 10.4 m Gran Telescopio CANARIAS. Non-sidereal guiding was not available at the time of the observations (it was fully implemented  in 2013, but later on) and therefore Apophis had to be tracked by applying offsets to the telescope every time the target was about to move off the CanariCam field of view (FOV) ($25'' \times 19''$). The standard star HD59381 was also observed on the same night as a flux calibrator. The data were taken in the Si2-8.7, Si6-12.5, and Q1-17.65 filters whose central wavelengths are 8.7, 12.5, and 17.65 $\mu$m, respectively (see Table~\ref{TableObs}). The telescope's secondary mirror was chopping at 2 Hz with a chop throw of $7''$\ along the east-west direction. Nodding of the telescope axes was performed every 47 seconds with a nod throw of $7''$, also in the east-west direction, to minimize the radiative offset.

\begin{table*}[h]
\caption{\label{TableObs}Log of observations.}
\centering
\begin{tabular}{lccclrr}
\hline\hline
Object   &    Date   &  UT start  &    UT end  &   Filter & on-source (s) & eff. on-source (s)  \\
\hline
Apophis & 2013-Jan-29 & 23:09:22.1 & 23:47:29.7 & Q1-17.65 & 908.316288 & 371.5839 \\
Apophis & 2013-Jan-29 & 23:52:38.6 & 23:56:13.0 & Si2-8.7  & 80.739226  & 80.73923 \\
Apophis & 2013-Jan-29 & 22:04:37.9 & 22:11:33.6 & Si6-12.5 & 165.148416 & 74.83288 \\
\hline
HD59381 & 2013-Jan-30 & 00:23:27.3 & 00:26:57.0 & Q1-17.65 & 82.574208  & 82.57421 \\
HD59381 & 2013-Jan-30 & 00:13:12.1 & 00:16:45.5 & Si2-8.7  & 80.739226  & 80.73923 \\
HD59381 & 2013-Jan-30 & 00:17:33.4 & 00:21:03.6 & Si6-12.5 & 82.574208  & 82.57421 \\
\hline
\end{tabular}
\tablefoot{Column (6) refers to the total on-source time of each observation. Column (7) refers to the actual on-source time used to create the final images, since in the case of Apophis not all savesets were usable to create the final images.}
\end{table*}

\begin{figure}
   \centering
   \includegraphics[width=6cm, angle=-90]{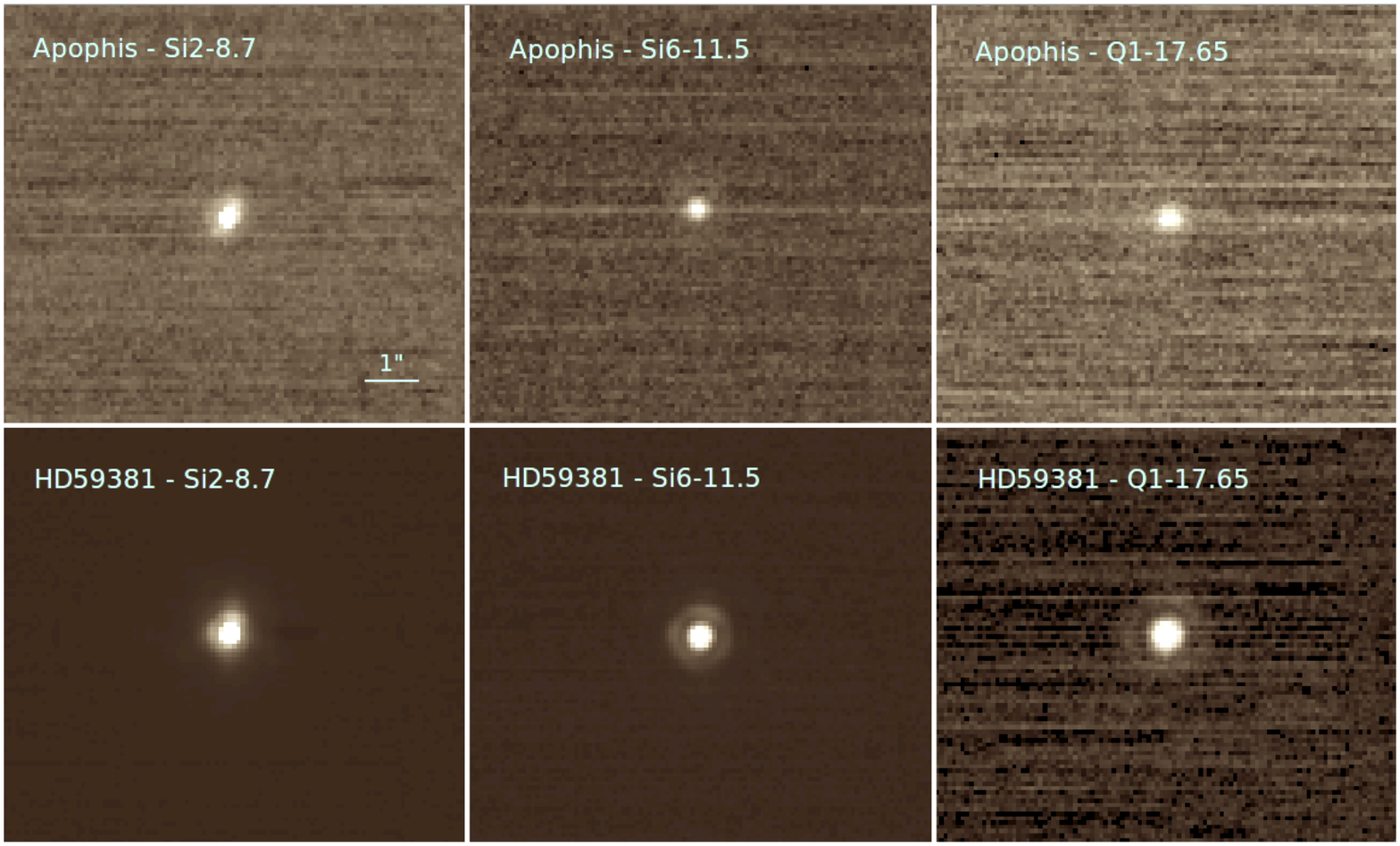}
      \caption{Composed images of Apophis (upper panels) and HD59381 flux standard star (lower panel) in the three  filters used.
}
         \label{apo_images}
   \end{figure}

Data were processed using a set of dedicated PyRAF~\footnote{PyRAF is a product of the Space Telescope Science Institute, which is operated by AURA for NASA.} scripts developed within our group. CanariCam raw images consist of a series of individual frames (savesets). The savesets are stored in multi-extension FITS files (MEF), which have the structure of [320,240,2,M][N]. The first two numbers represent the detector's X and Y dimensions in pixels ($320 \times 240$). The third dimension represents the number of chop positions, namely on-source and off-source positions. M represents the number of savesets in each nod position and N the number of nods (nod beams A and B), which follow the sequence A-BB-A. Off-source savesets were subtracted from the corresponding on-source savesets for each nod beam. For each individual saveset, we determined the source centroid using SExtractor (Bertin \& Arnouts \cite{BertinArnouts}). Savesets were then geometrically aligned using the shifts calculated from the centroid positions with respect to centroid from the first saveset and stacked to produce the final net signal image. This shift-and-add technique improves the image quality, and therefore the sensitivity. Shift-and-add is particularly important for the Apophis data since the object was drifting accross the CanariCam detector by several pixels from one saveset to the next. Additionally, owing to the relatively short on-source time in each saveset (5.9 s) and the rapid movement of Apophis accross the FOV, it was not possible to calculate image centroid in all savesets. Hence, those savesets where the centroid was not obtained were discarded from the total signal, yielding an effective on-source time smaller than the observed on-source time (see Table \ref{TablePhot}). The resulting images of Apophis and the standard star used are shown in Figure~\ref{apo_images}.

Aperture photometry was performed in the reduced images using PyRAF. An aperture of radius $0.6''$\ was used in all filters in the Apophis as well as in the standard star HD59381 images. A sky annulus with a radius of $2.4''$\ and  width of $0.8''$\ was used to determine the sky as the median of all pixel values within the annulus area. The in-band flux in Jy for each CanariCam filter was obtained by integrating the standard star template spectrum (Cohen et al. \cite{Cohen}) multiplied by the filter transmission curve. Finally, the in-band flux for HD59381 was divided by the measured ADU/s within the $0.6''$\ radius aperture and then multiplied by the measured ADU/s within the same aperture in Apophis. No color correction is applied as it is much smaller ($< 1\%$) than the uncertainties. The final flux densities and the FWHM of the PSF in each image can be found in Table~\ref{TablePhot}.

\begin{table}[h]
\caption{\label{TablePhot}Photometric data.}
\centering
\begin{tabular}{lccc}
\hline\hline
Object  &  Filter  &      Flux (Jy)     &    FWHM ($''$)  \\
\hline
Apophis & Si2-8.7  & $0.14 \pm 0.01 $ & $0.25 \pm 0.01$ \\
Apophis & Si6-12.5 & $0.24 \pm 0.02$    & $0.32 \pm 0.04$ \\
Apophis & Q1-17.65 & $0.31 \pm 0.07$    & $0.43 \pm 0.01$ \\
\hline
HD59381 & Si2-8.7  & $8.80 \pm 0.04$      & $0.25 \pm 0.01$ \\
HD59381 & Si6-12.5 & $4.93 \pm 0.04$      & $0.32 \pm 0.04$ \\
HD59381 & Q1-17.65 & $2.43 \pm 0.04$      & $0.43 \pm 0.01$ \\
\hline
\end{tabular}
\tablefoot{Apophis fluxes correspond to an aperture of radius $0.6''$. HD59381 fluxes were obtained by integrating the Cohen et al. \cite{Cohen} template spectrum multiplied by the CanariCam filter transmission curves.}
\end{table}

\section{Thermophysical modeling}\label{sec:TPM}
Thermophysical models (TPM) are powerful tools used to derive asteroid sizes from their thermal infrared data. If the shape, the spin axis orientation, and the rotational period of the object are well characterized and if enough data are available, TPMs also allow  the thermal inertia and the macroscopic roughness of the surface  to be constrained. 
In brief, TPMs model the temperature on each surface element of the shape model --typically triangular facets-- at every observation epoch by accounting for the corresponding energy budget, i.e., how much incident solar radiation is absorbed at and conducted onto the surface. 
This depends on the distance from the asteroid to the Sun and on the physical properties of the surface (albedo, emissivity, macroscopic roughness, conductivity, etc.), but it also depends critically on the shape of the object and its rotational phase at the moment of the observations since they determine the illumination geometry of each facet. 
Thus, the TPM requires any available convex shape model in combination with the spin axis orientation and rotational properties as input.

The heat conduction onto the surface is controlled by the thermal inertia $\Gamma$, while the infrared beaming effects are calculated via a surface roughness model implemented as concave, spherical crater segments on the surface and parametrized by the root mean square ({\it rms}) slope angle.
Once the temperatures are modeled,  the model fluxes that the observer would measure can be computed given the particular observational geometry and they can be fit to the data.
The observational geometry refers to the heliocentric and geocentric distances and the phase angle --the angle subtended by the observer and the Sun from the point of view of the asteroid.

\begin{figure}
   \centering
   \includegraphics[width=7cm, angle=90]{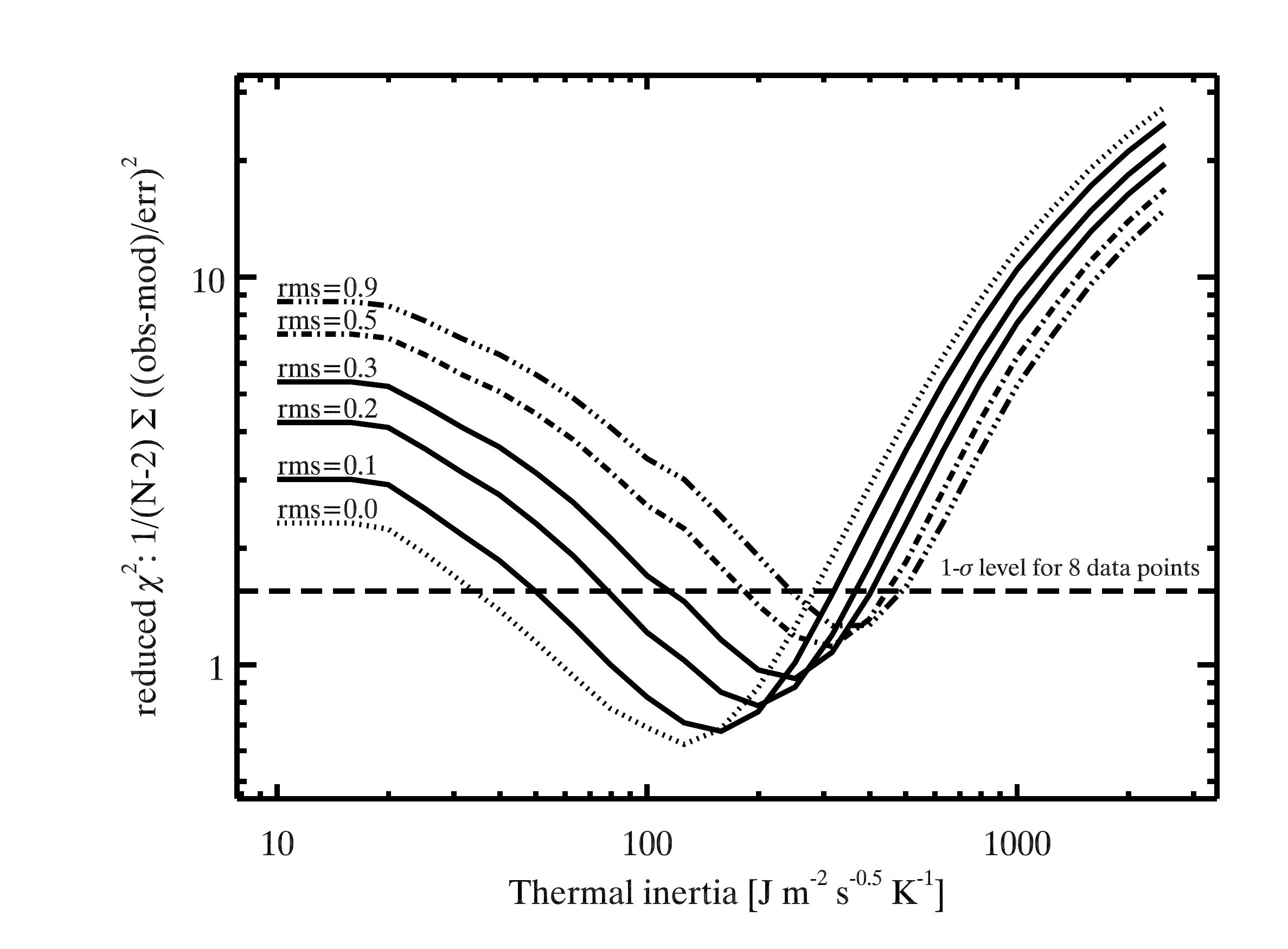}
      \caption{Reduced $\chi^2$ values for models considering different roughness slope angles $rms =$ 0.0, 0.1, 0.2, 0.3, 0.5 and 0.9. Notice that the minima of the reduced $\chi^2$-values are lower for low-roughness solutions, but we find acceptable fits (with  $\chi^2$-values lower than 1.55 shown as an horizontal line) for all levels of roughness. 
 }
         \label{chi2}
   \end{figure}

M\"uller et al. (2014) applied a TPM based on the work by Lagerros (\cite{Lagerros96}, \cite{Lagerros97}, \cite{Lagerros98}) and M\"uller \& Lagerros (\cite{Mueller1998}, \cite{Mueller2002}) to model the thermal data of Apophis obtained with  the {\em Herschel Space Observatory} PACS instrument. 
Taking the tumbling rotational state and shape model given by Pravec et al. (\cite{Pravec2014}), M\"uller et al. obtained a thermal inertia of $\Gamma = 600 ^{+200}_{-350}$ Jm$^{-2}$s$^{-0.5}$K$^{-1}$, an effective diameter of $D_{eff} = 375 ^{+14}_{-10}$ m, and a corresponding visible ($V$-band) geometric albedo of $p_V = 0.30 ^{+0.05}_{-0.06}$ for Apophis. 
In this paper we use the same TPM and rotational state model to study a combination of GTC and {\em Herschel} data to better constrain the M\"uller et al. (2014) results.  
We combined all GTC and {\em Herschel} data using the results obtained in M\"uller et al. (\cite{Mueller2014}), with {\em Herschel} data alone as a starting guess. We assumed a constant emissivity of 0.9 at all wavelengths. 
We also used the mean absolute magnitude $H_V = 19.09 \pm 0.19$ mag derived by Pravec et al. (\cite{Pravec2014}) under the assumption of a slope parameter of $G = 0.24 \pm 0.11$.   The observational circumstances of {\em Herschel Space Observatory} data are summarized in Table 1 of  M\"uller et al. (\cite{Mueller2014}); GTC observations were obtained with the asteroid at heliocentric and geocentric distances r$_{helio} = 1.080$ AU and  $\Delta_{obs} = 0.113,$ respectively, and a phase angle $\alpha = -31.7$ degrees. The illumination is similar to that  shown in M\"uller et al. (\cite{Mueller2014}), Fig. 3, right panel, except that it is at a smaller phase angle (CanariCam: -31.7 deg, PACS: -61.4 deg) and a different orientation of the body (as seen from GTC).
In Fig. \ref{chi2} we plot the reduced $\chi^2$-values calculated for the radiometric analysis of the combined GTC and {\em Herschel} data for different roughness slope angles versus thermal inertia. 
The roughness slope angles ($rms$) range from 0.0 to 0.9.

\begin{figure}
   \centering
   \includegraphics[width=6cm, angle=90]{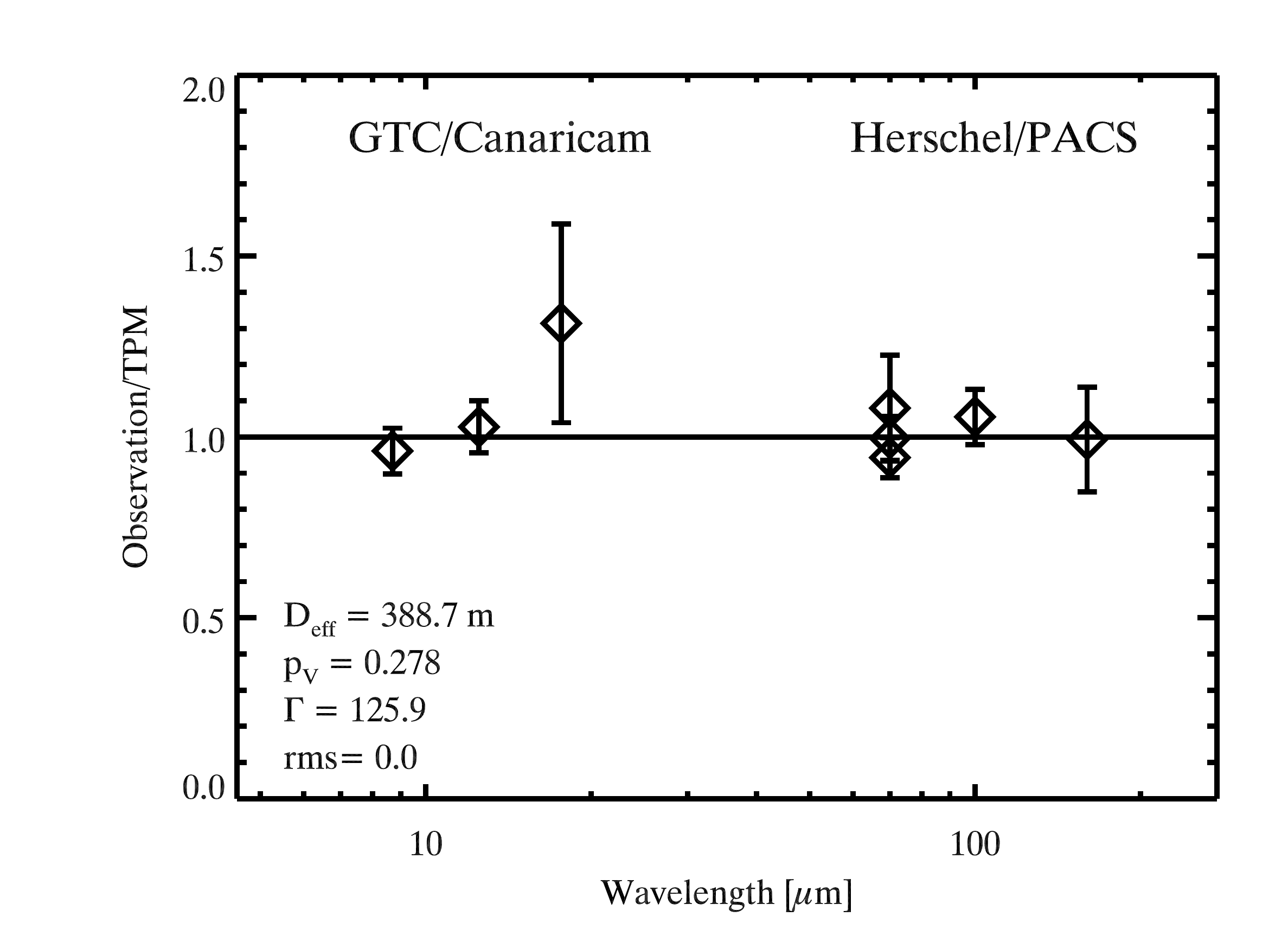}
\includegraphics[width=6cm, angle=90]{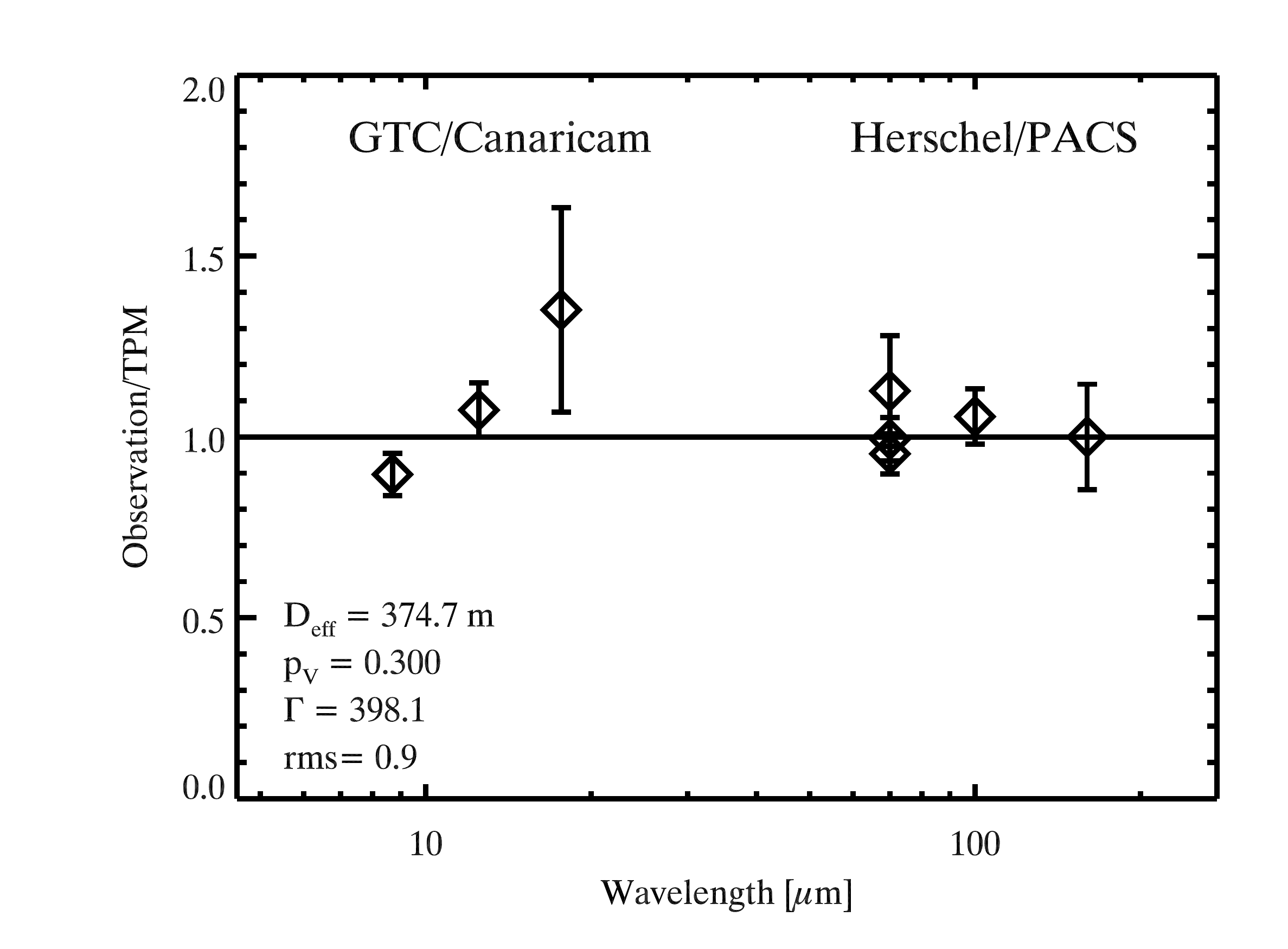}\\
      \caption{Observed GTC/CanariCam (this paper) and Herschel/PACS (from M\"uller et al. \cite{Mueller2014}) divided by model fluxes for the two extreme cases with roughness slope angles $rms =$ 0.0 ($upper$) and 0.9 ($lower$). Notice that observed-to-modeled fluxes of the GTC/CanariCam data are particularly sensitive to roughness and worsen for larger roughness, which led us to favor the solutions with low surface roughness in our analysis}
         \label{ajustes}
   \end{figure}

The minimum reduced  $\chi^2$ would be $\sim$0.8 and the statistical error would be $\sigma\sim 0.6$ so we find acceptable fits (with $\chi^2$-values lower than 1.55) for all levels of roughness (see, e.g., Press et al. \cite{Press1986}), which means that we cannot unambiguously constrain roughness and thermal inertia: low-roughness combined with low $\Gamma$ fit similarly well to high-roughness and high-$\Gamma$ solutions. 
On the other hand, the minima of the reduced $\chi^2$-values are lower for low-roughness solutions. This is also illustrated in the ratios of observed-to-modeled fluxes presented in Fig. \ref{ajustes}, where we compare the ratios of the observed fluxes to our modeled fluxes for the extreme roughness cases. We note in particular that observed-to-modeled fluxes of the GTC/CanariCam data  are particularly sensitive to roughness and are a slightly less sensitive for larger roughness. These considerations led us to favor the solutions with low surface roughness in our analysis. Thus, within a range of $rms$ between 0.1 and 0.5, we constrain the size, visible geometric albedo, and thermal inertia of Apophis to be $D_{eff} =$~380 -- 393 m, $p_V = $~0.27--0.29, and $\Gamma =$~50 -- 500 Jm$^{-2}$s$^{-0.5}$K$^{-1}$. We note that the given range of $p_V$ is obtained assuming the absolute magnitude $H=19.09$ from Pravec et al. (\cite{Pravec2014}), considering the uncertainty in $H$ ($H = 19.09 \pm 0.19$) and so the range of possible albedo is wider ($p_V = $~0.24--0.33).

\begin{table}[h]
\caption{\label{Results}  $D_{eff} $, $p_V$, and $\Gamma$ values corresponding to the minimum $\chi^2$ value of the TPM fits using fixed values of $rms$ in the 0.0 to 0.9 range.}
\centering
\begin{tabular}{ccccc}
\hline\hline
rms & D$_{eff}$ (m) & p$_V$  & $\Gamma$ & $\chi^2_{reduced}$ \\
\noalign{\smallskip} \hline \noalign{\smallskip}
0.0 & 389           & 0.278  & 126      & 0.624 \\
0.1 & 389           & 0.278  & 159      & 0.674 \\
0.2 & 387           & 0.281  & 200      & 0.784 \\
0.3 & 384           & 0.285  & 251      & 0.922 \\
0.5 & 380           & 0.292  & 316      & 1.114 \\
0.9 & 375           & 0.300  & 398      & 1.262 \\
\noalign{\smallskip} \hline \noalign{\smallskip}
\hline
\end{tabular}
\end{table}

\section{Discussion and conclusions}\label{sec:discussion}

Images of Apophis were obtained using three different filters (Si2-8.7, Si6-12.5, and Q1-17.65) with the CanariCam instrument in imaging mode at the 10.4 m Gran Telescopio CANARIAS (GTC) at El Roque de los Muchachos Observatory (La Palma, Canary Islands, Spain). The derived fluxes, which are closer to the wavelengths in which Apophis' thermal emission peaks than those reported by M\"uller et al. (\cite{Mueller2014}) using {\it Herschel} space telescope data, are used together with M\"uller et al. reported fluxes to better constrain the thermophysical model of Apophis also presented in M\"uller et al. 

Our fitting of the TPM to the combined GTC/CanariCam and PACS/$Herschel$  favors the solutions with low surface roughness  (within a range of  $rms$ roughness slope angles between 0.1 and 0.5) and it constrains the size, visible geometric albedo, and thermal inertia of Apophis to be $D_{eff} =$~380 -- 393 m, $p_V = $~0.27--0.29, and $\Gamma =$~50 -- 500 Jm$^{-2}$ s$^{-0.5}$ K$^{-1}$. 

These results agree very well within the uncertainties with those reported by M\"uller et al. (\cite{Mueller2014})  using only the $Herschel$ data   ($D_{eff} = 375^{+14}_{-10}$ m, $p_V = 0.30^{+0.05}_{-0.06}$, a thermal inertia $\Gamma $ in the range 250-800 Jm$^{-2}$ s$^{-0.5}$ K$^{-1}$, with a best solution at $\Gamma =$~600 Jm$^{-2}$ s$^{-0.5}$ K$^{-1}$ for  $rms = 0.5$), but point  to a somewhat lower value of thermal inertia, closer to the lower range given in M\"uller et al., and a slightly larger effective size (closer to the wider range given in M\"uller et al.). The albedo value also agrees with the value reported by Delbo et al. (\cite{Delbo2007}) ($p_V =0.33 \pm 0.08$) derived from polarimetric observations. The thermal inertia of Apophis also closely fits   the trend of $\Gamma$ vs. $D_{eff}$ reported by Delbo \& Tanga (\cite{Delbo2009}). 

Finally, the presented improvement in the determination of the thermal inertia is very important in order to evaluate the Yarkovsky effect on the orbital evolution of Apophis (Vokrouhlick\'y et al., \cite{Vokrouhlicky}). Using the M\"uller et al. (\cite{Mueller2014})  range of possible $\Gamma$-values, they estimate the drift of the Apophis orbital semimajor axis $\langle da/dt \rangle$  to be in a range between $-11$  x $10^{-4}$ and $-15$ x $10^{-4}$ au/Myear.  Using the range of thermal inertia obtained in this paper, the range of $\langle da/dt \rangle$ that can be derived from  Fig. 1 in Vokrouhlick\'y et al. (\cite{Vokrouhlicky}) is slightly smaller, between $-6$  x $10^{-4}$ and $-14$ x $10^{-4}$ au/Myear.



\begin{acknowledgements}
We acknowledge Ben Rositis for his useful comments that helped to improve the manuscript. Based on observations made with the Gran Telescopio Canarias (GTC), installed in the Spanish Observatorio del Roque de los Muchachos of the Instituto de Astrof\'{\i}sica de Canarias, in the island of La Palma.

We acknowledge Petr Scheirich \& Petr Pravec for the provision of  shape orientation, spin-axis orientation, rotational
properties, etc for the GTC observing epochs, and Ben Rositis for his useful comments that helped to improve the manuscript. JL acknowledges support from the project ESP2013-47816-C4-2-P (MINECO, Spanish Ministry of Economy and Competitiveness). MD and V. Ali-Lagoa acknowledge support from the NEOShield-2 project that has received funding from the European UnionÕs Horizon 2020 research and innovation program under grant agreement No 640351.
The work of V. Ali-Lagoa and MDB was supported by the French Agence National de la Recherche (ANR) SHOCKS

\end{acknowledgements}



\begin{thebibliography}{99}

\bibitem[1996]{BertinArnouts} 
Bertin, E. \& Arnouts, S., 1996, A\&AS, 117, 393

\bibitem[2002]{Botkke2002} 
Bottke, W. F., Morbidelli, A., Jedicke, R. et al., 2002, Icarus, 156, 399
\bibitem[1999]{Cohen} 
Cohen, M.,  Walker, R. G. Carter, B. et al. 1999, AJ, 117, 1864

\bibitem[2009]{Delbo2009} 
Delbo, M.,\& Tanga, P., 2009, P\&SS, 57, 259

\bibitem[2007]{Delbo2007} 
Delbo, M., Cellino, A. \& Tedesco, E., 2007, Icarus, 188, 266

\bibitem[2013]{Farnocchia2013} 
Farnocchia, D., Chesley, S. R., Chodas, P.W., et al. 2013, Icarus, 224, 192


\bibitem[2002]{Giorgini2002}
Giorgini, J., Ostro, J., Benner, L. et al., 2002, Science, 296,132

\bibitem[1996]{Lagerros96} 
Lagerros, J. S. V. 1996, A\&A, 310, 1011

\bibitem[1997]{Lagerros97} 
Lagerros, J. S. V. 1997, A\&A, 325, 1226

\bibitem[1998]{Lagerros98} 
Lagerros, J. S. V. 1998, A\&A, 332, 1123

\bibitem[1998]{Mueller1998} 
M\"uller, T. \& Lagerros, J. S. V. 1998,  A\&A, 338, 340

\bibitem[2002]{Mueller2002} 
M\"uller, T. \& Lagerros, J. S. V. 2002, A\&A, 381, 324

\bibitem[2014]{Mueller2014} 
M\"uller, T., Kiss, C., Scheirich, P. et al., 2014, A\&A, 566, 22M

\bibitem[2014]{Pravec2014} 
Pravec, P., Scheirich, P., Durech, J. et al. 2014, Icarus, 233, 48

\bibitem[1986]{Press1986} 
Press, W. H., Flannery, B. P., \& Teukolsky, S. A. 1986, Numerical recipes (Cambridge: University Press)

\bibitem[2003]{Telesco} 
Telesco C. M., Ciardi D., French J., et al., 2003, in SPIE Conference Series, Vol. 4841, 913

\bibitem[2015]{Vokrouhlicky} 
Vokrouhlick\'y, D., Farnocchia, D., Capek, D. et al. 2015, Icarus, 252, 277


\end{thebibliography}
\end{document}